\documentclass[runningheads]{llncs}
\usepackage[T1]{fontenc}
\usepackage{graphicx}
\usepackage{graphicx}
\usepackage{textcomp}
\usepackage{xcolor}
\usepackage{amsmath, calc}
\usepackage{tcolorbox}
\usepackage{multirow}
\usepackage{url}
\usepackage{hyperref}
\usepackage{pgfplots}
\pgfplotsset{compat=1.18}
\usepackage{placeins}
\usepackage{fontawesome}
\usepackage{comment}
\usepackage{tabularx}
\usepackage{booktabs}
\usepackage{framed}
\usepackage{fancybox}
\usepackage{makecell}
\usepackage{csquotes}
\usepackage{orcidlink}
\usepackage{multicol}
\usepackage{colortbl}
\usepackage{caption}
\usepackage{amssymb}

\begin{document}
\title{Understanding Usefulness in Developer Explanations on Stack Overflow}
\titlerunning{Understanding Usefulness in Developer Explanations on Stack Overflow}

%
\author{
    Martin Obaidi\orcidlink{0000-0001-9217-3934}\inst{1} \and
    Kushtrim Qengaj\inst{1} \and
    Hannah Deters\orcidlink{0000-0001-9077-7486}\inst{1} \and
    Jakob Droste\orcidlink{0000-0001-8746-6329}\inst{1} \and
    Marc Herrmann\orcidlink{0000-0002-3951-3300}\inst{1} \and
    Kurt Schneider\orcidlink{0000-0002-7456-8323}\inst{1} \and
    Jil Klünder\orcidlink{0000-0001-7674-2930}\inst{2}
}

\authorrunning{Obaidi et al.} 

\institute{
    Leibniz Universität Hannover, Software Engineering Group, Hannover, Germany \\
    \email{\{martin.obaidi, hannah.deters, jakob.droste, marc.herrmann, kurt.schneider\}@inf.uni-hannover.de} \\
    \email{kushtrim.qengaj@stud.uni-hannover.de}
    \and
    University of Applied Sciences | FHDW Hannover, Germany \\
    \email{jil.kluender@fhdw.de}
}
\maketitle              

\begin{abstract}
\textbf{[Context and motivation]}  
Explanations are essential in software engineering (SE) and requirements communication, helping stakeholders clarify ambiguities, justify design choices, and build shared understanding. Online Q\&A forums such as \textit{Stack Overflow} provide large-scale settings where such explanations are produced and evaluated, offering valuable insights into what makes them effective.

\textbf{[Question/problem]}  
While prior work has explored answer acceptance and voting behavior, little is known about which specific features make explanations genuinely useful. The relative influence of structural, contextual, and linguistic factors, such as content richness, timing, and sentiment, remains unclear.

\textbf{[Principal ideas/results]}  
We analyzed 3,323 questions and 59,398 answers from \textit{Stack Overflow}, combining text analysis and statistical modeling to examine how explanation attributes relate to perceived usefulness (normalized upvotes). Structural and contextual factors, especially explanation length, code inclusion, timing, and author reputation, show small to moderate positive effects. Sentiment polarity has negligible influence, suggesting that clarity and substance outweigh tone in technical communication.

\textbf{[Contribution]}  
This study provides an empirical account of what drives perceived usefulness in developer explanations. It contributes methodological transparency through open data and replication materials, and conceptual insight by relating observed communication patterns to principles of requirements communication. The findings offer evidence-based implications for how developers and RE practitioners can craft clearer and more effective explanations, potentially supporting fairer communication in both open and organizational contexts. From an RE perspective, these determinants can be interpreted as practical signals for ambiguity reduction and rationale articulation in day-to-day requirements communication.

\end{abstract}

\section{Introduction}
\label{sec:intro}

With the rising complexity of software systems, particularly in data-driven and requirements-intensive domains, clear and comprehensible explanations are becoming ever more essential~\cite{adadi2018peeking,deters2024UXandExplainability}.  In requirements engineering (RE), explanations help clarify requirements, resolve ambiguities, and support shared understanding among stakeholders~\cite{chazette2021exploring,droste2024explanations}. Clarity and shared understanding have long been linked to successful RE practice~\cite{hofmann2001requirements}. Yet, producing high-quality, context-sensitive explanations remains difficult as artifacts, tools, and teams scale. In practice, breakdowns in RE often stem from missing or weakly communicated rationale and delayed clarification, which can propagate ambiguity into implementation and validation. 
Thus, understanding which explanation attributes developers perceive as useful provides an empirical basis for improving RE communication practices that aim at shared understanding and ambiguity reduction.

Building on the definition of explainability by Chazette et al.~\cite{chazette2021exploring}, we focus on \textit{developer explainability}—how developers explain software behaviour, code, or design decisions to peers. Such explanations act as coordination mechanisms that align mental models and reduce uncertainty~\cite{chandra2024watchatexplainingperplexingprograms}. Recent work further shows that context-proximate explanations embedded in development workflows can improve comprehension and confidence~\cite{yan2024ExplainDeveloper}. 
In software projects, the ability to explain and understand technical or design rationales directly affects how requirements are interpreted and implemented. Satisfying explanatory needs among developers therefore helps preserve requirement intent and ensures that implementation aligns with stakeholder expectations.
In RE terms, these explainability practices parallel activities such as rationale articulation and clarification during elicitation and validation, where stakeholders exchange explanations to achieve mutual understanding.

Online Q\&A platforms such as \textit{Stack Overflow} provide large-scale evidence of this explanatory behaviour. Prior work has characterized \textit{Stack Overflow} as particularly effective for conceptual questions and code-review-like exchanges, noting that review-oriented questions are often concrete and include code snippets that facilitate explanation and assessment~\cite{treude2011askquestions}. Building on this foundation, our study quantifies how specific structural, contextual, and linguistic attributes of explanations relate to \emph{perceived usefulness} at scale. Engagement signals like upvotes reflect community judgments of usefulness, albeit biased by visibility, timing, and author reputation~\cite{anderson2012stackoverflow,mondal2023SOSubjectivity}. Prior studies have explored factors influencing answer quality, such as code inclusion, length, and response time~\cite{calefato2015miningSO,omondiagbe2019features}, yet little is known about which structural, contextual, or linguistic features make explanations themselves \textit{useful} in RE-related discussions.

To address this gap, we empirically analyze 3{,}323 questions and 59{,}398 answers from \textit{Stack Overflow} to examine how structural (e.g., content richness), contextual (e.g., timing, reputation), and linguistic (e.g., sentiment) factors relate to perceived usefulness. Specifically, we investigate which characteristics distinguish explanations that developers perceive as more useful in technical and RE-oriented communication.
We conceptualize developer explainability as a communicative competence that supports rationale sharing and ambiguity reduction, which are key qualities of effective RE communication. 
By identifying key drivers of perceived usefulness, this study contributes to a better understanding of how developers communicate explanations effectively. The insights aim to inform RE communication practices and the design of tools that foster clear, fair, and context-aware explanations. The remainder of this paper presents related work (Section~\ref{sec:related}), the study design (Section~\ref{sec:research}), results (Section~\ref{sec:results}), discussion (Section~\ref{sec:discussion}), and conclusion (Section~\ref{sec:conclusion}).

\section{Background and Related Work}
\label{sec:related}

\subsection{Background}

Explainability is a software quality aspect aimed at making systems and their behavior understandable to humans~\cite{chazette2021exploring}. 
Beyond its roots in AI transparency~\cite{adadi2018peeking}, explainability has become relevant across software engineering (SE), where understanding and rationale exchange are essential for effective collaboration~\cite{droste2024explanations}. 
Following Chazette et al.~\cite{chazette2021exploring}, we view explainability as a non-functional requirement (NFR) closely linked to transparency, trust, and usability. 

We extend this concept toward \textit{developer explainability}—how developers explain technical artifacts such as code, APIs, or design decisions to peers. 
Explanations in this sense facilitate understanding, reduce ambiguity, and support shared reasoning, making them critical for both development and requirements communication. 
When well designed, they enhance clarity and trust; when poor or excessive, they may hinder comprehension~\cite{deters2024UXandExplainability}. 
On-demand environments such as \textit{Stack Overflow} exemplify this dynamic by enabling developers to request and evaluate explanations in real time~\cite{gupta2016stackoverflow}.

\subsection{Related Work}

\subsubsection{Explainability in Software Engineering}
Recent work increasingly investigates explainability as a NFR beyond XAI~\cite{obaidi2025appKonwledge,obaidi2025automatingexplanationneedmanagement,obaidi2025elicit,obaidi2025mood,deters2024qualitymodel,unterbusch2023explanation,droste2025operationaltaxonomy,deters2025explanationcatalog,deters2025metrics}.
Chazette et al.~\cite{chazette2021exploring} conceptualized explainability as an NFR and highlighted its context-dependent and potentially conflicting effects. 
Building on this, Chandra et al.~\cite{chandra2024watchatexplainingperplexingprograms} and Yan et al.~\cite{yan2024ExplainDeveloper} showed that embedded, in-situ explanations improve comprehension and help align developers’ mental models. 
Research on developer communication~\cite{alkadhi2017developmentChat,chatterjee2021qualityDeveloperChats} further indicates that a large share of developer dialogue involves rationale sharing and explanation construction, suggesting that explainability is not confined to user-facing AI but pervades everyday collaboration.

\subsubsection{Explanation Usefulness on Q\&A Platforms}

Online Q\&A platforms such as \textit{Stack Overflow} provide large-scale data on peer-generated explanations. 
Prior work links explanation quality to structural and contextual factors: code inclusion, textual length, and timing~\cite{nasehi2012goodExample,omondiagbe2019features,calefato2015miningSO,bhat2015timeSO}. 
User reputation and engagement also correlate with perceived quality~\cite{PROCACI2015664,wang2021reputation}. 
However, popularity does not always align with accuracy or clarity~\cite{mondal2023SOSubjectivity}. 
Few studies explicitly interpret these determinants as facets of \textit{developer explainability}—how explanations convey rationale and foster understanding within technical discussions. 

Our study extends this perspective by analyzing structural, temporal, and social factors as predictors of explanation usefulness, linking insights from Q\&A research to explainability and requirements communication.

\subsubsection{Sentiment Analysis}

Sentiment analysis in SE has gained significant attention over the past decade as a means to assess affective tone in developer communication such as discussions or and Q\&A posts~\cite{zhang2025sentimentllm,cassee2024sentimentSE,obaidiSentiSMS22,obaidi2021development,obaidi2025GoldStandardDE,herrmann2025montecarlo,herrmann2025different-perceptions,schroth2022potential,herrmannSentiSurvey22}. Calefato et al.~\cite{calefato2018senti} introduced \textit{Senti4SD}, a supervised classifier trained on over 4,400 \textit{Stack Overflow} posts, demonstrating that SE-specific sentiment models outperform generic social-media tools. 
Novielli et al.~\cite{novielli2018stackoverflow-gold} released a manually annotated gold-standard dataset of 4,800 Stack Overflow posts, enabling benchmarking and reproducibility across sentiment studies. 
Swillus and Zaidman~\cite{swillus2023sentiment} applied sentiment analysis to testing-related Stack Overflow posts, uncovering affective dimensions such as insecurity or aspiration that influence attitudes toward testing practice. 
Complementarily, Uddin et al.~\cite{uddin2021SO} focused on opinion mining in API discussions, automatically classifying sentiments toward API usability and performance aspects. Recent work~\cite{zhang2025sentimentllm,obaidi2025TrustworthySA,obaidi22cross} further demonstrates that fine-tuned transformer models (e.g., BERT, RoBERTa) achieve state-of-the-art performance for polarity detection in software-related text.

However, prior work primarily associates sentiment with emotional or experiential factors—such as developer satisfaction, motivation, or team climate—rather than with the perceived usefulness of content. 
Our study addresses this gap by examining whether the affective tone of explanations has any measurable effect on their perceived usefulness within technical discussions.

\section{Study Design}
\label{sec:research}

\begin{figure*}[htbp]
    \centering
    \includegraphics[width=0.8\linewidth]{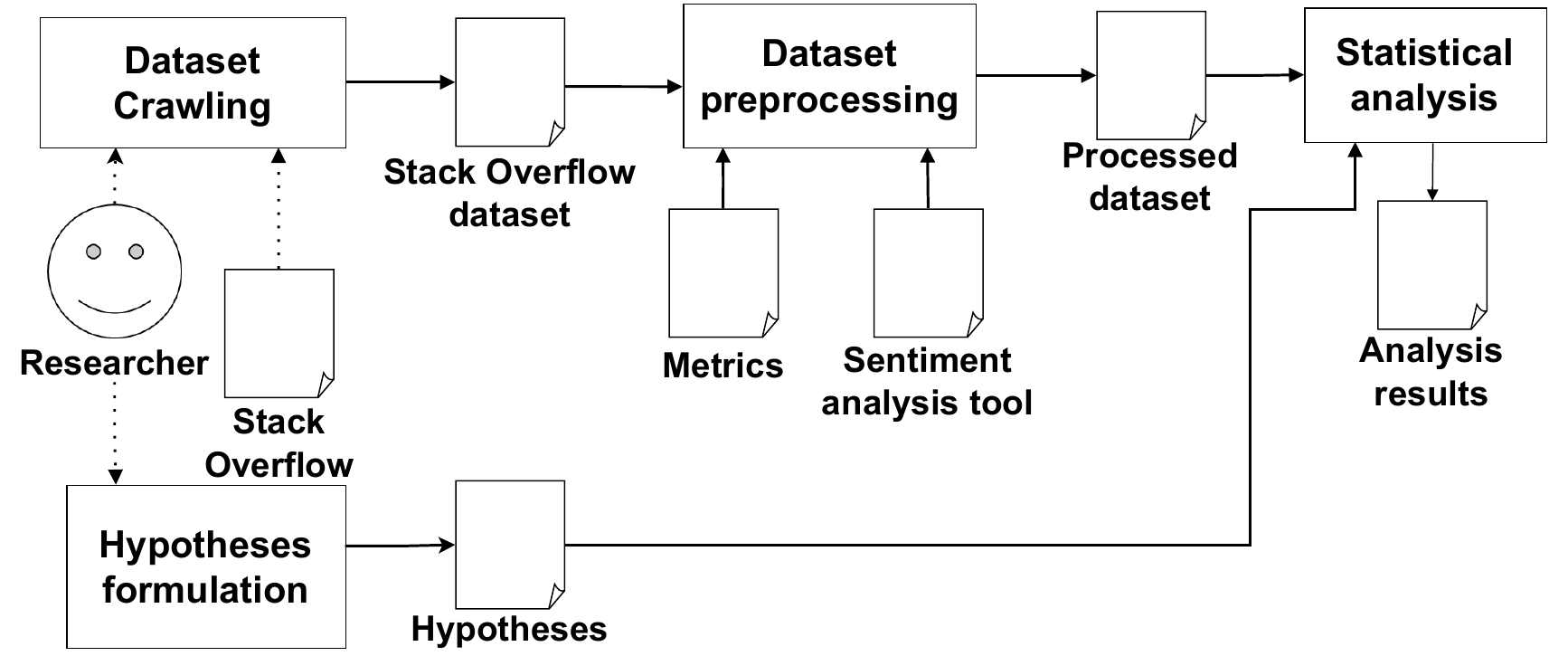}
    \caption{Overview of the study workflow following FLOW notation~\cite{stapel2009flow}. The symbols are chosen accordingly to the notation~\cite{stapel2009flow}.}
    \label{fig:studie}
\end{figure*}

This study investigates which structural, contextual, and social factors contribute to the perceived usefulness of developer explanations on \textit{Stack Overflow}. 
Following the workflow shown in Figure~\ref{fig:studie}, we collected and preprocessed data, derived explanatory metrics, and applied correlation analyses to identify significant associations.

Methodologically, our metadata-driven, hypothesis-based correlation approach is in line with prior explainability research that linked explanation-related constructs to platform metadata in other domains (e.g., app reviews)~\cite{obaidi2025AppFeaturesExplainNeeds}.

\subsection{Research Questions}

The goal is to identify what makes explanations in developer discussions useful and how this relates to \textit{developer explainability}. 
We address four research questions:

\begin{itemize}
    \item \textbf{RQ1:} How does \emph{perceived usefulness} relate to the sentiment polarity of an explanation?
    \item \textbf{RQ2:} How does \emph{perceived usefulness} relate to content features (e.g., text length or readability)?
    \item \textbf{RQ3:} How does \emph{perceived usefulness} relate to timing and visibility factors (e.g., posting delay and response order)?
    \item \textbf{RQ4:} How does \emph{perceived usefulness} relate to author characteristics (e.g., reputation and badges)?
\end{itemize}

These questions capture linguistic, structural, temporal, and social determinants of explanation usefulness, extending prior Q\&A research toward RE-related communication.

\subsection{Dataset and Forum Selection}

\textit{Stack Overflow} was selected for its structured Q\&A format and rich metadata (scores, timestamps, user reputation). 
We focused on posts tagged \texttt{[android]} because it is a long-running and high-volume tag with comparatively stable usage over time, and it contains a balanced mix of conceptual questions (e.g., design and API usage) and code-level debugging discussions, providing sufficient variation and sample size for robust statistical estimation. 
To ensure high-quality data, only questions with at least 50 upvotes and at least one positively rated answer were included. 
This yielded 3,323 questions and 59,398 answers posted between 2009 and 2024. 
Duplicate and incomplete entries from API exports were removed manually. 
We treat this dataset as a case study of a large developer community. Replication across additional tags and platforms is needed to assess generalizability.

These thresholds were chosen to ensure representativeness and linguistic substance: highly upvoted questions better reflect community consensus, while a minimum word count of 100 (see Section~\ref{subsec:data-preprocessing}) ensures textual completeness for reliable linguistic analysis.

\subsection{Metrics and Operationalization}

Perceived usefulness was approximated through an answer’s \textit{relative score}, i.e., its score normalized by the total score of all answers to the same question. This allows comparisons across threads of differing popularity. 
Importantly, votes and scores should be interpreted as a noisy signal of \emph{perceived} usefulness in a specific platform context: they may reflect explanatory value, but are also influenced by visibility, timing, author reputation, and community norms, and therefore do not directly measure correctness or understanding.

Independent variables were grouped as follows:

\begin{itemize}
    \item \textbf{Sentiment:} sentiment polarity (negative, neutral, positive) obtained via a fine-tuned transformer model~\cite{Devlin.2019bert} applied to the cleaned answer text.
    \item \textbf{Content:} number of code blocks, links, paragraphs, sentences, and words; readability (Flesch Reading Ease); lexical diversity; and textual similarity to the question and other answers (Jaccard coefficient over lemmatized tokens).
    \item \textbf{Timing:} creation date, time since the corresponding question, and posting order.
    \item \textbf{Author:} reputation and badge counts (bronze, silver, gold) of the answer owner and last editor.
\end{itemize}

All variables were scaled or log-transformed where appropriate to ensure comparability. 
Quantitative metrics were computed automatically using \textit{textstat} and \textit{LexicalRichness}; code and link counts were extracted via regular expressions.

\subsection{Data Preprocessing}
\label{subsec:data-preprocessing}

For each answer, raw and processed text versions were retained. 
Preprocessing involved removing code blocks and HTML, decoding entities, eliminating stop words, and lemmatizing tokens~\cite{schutze2008introduction}. 
Answers shorter than 100 words were excluded from linguistic analyses (e.g., readability, similarity) to maintain statistical reliability~\cite{graesser2004cohmetrix}. 
All preprocessing was performed using \textit{BeautifulSoup~4.12}, \textit{NLTK~3.9}, and \textit{spaCy~3.7} (\textit{en\_core\_web\_sm} model), ensuring consistent tokenization and stopword handling.

\subsection{Analysis Procedure}

We conducted correlation analyses to test associations between variables and usefulness, selecting the coefficient according to the measurement scale of each predictor: Spearman’s $\rho$ for ordinal/continuous variables, point-biserial $r_{pb}$ for binary variables, and eta $\eta$ for nominal groupings. These coefficients are used as descriptive association measures and are not contingent on normally distributed variables. Accordingly, we focus on effect sizes with Bonferroni-corrected thresholds for multiple testing. This design follows the same general rationale as metadata-based explainability studies that relate observable platform signals to explanation phenomena~\cite{obaidi2025AppFeaturesExplainNeeds}.

\subsection{Data Analysis}

All analyses were conducted using established scientific libraries (\textit{pandas}, \textit{numpy}, \textit{scipy}, \textit{pingouin}, \textit{statsmodels}). 
This ensured reproducibility and statistical comparability across all analytical steps.

\subsubsection{Sentiment Analysis}
\label{subsec:sentiment-analysis}

To estimate the emotional tone of explanations, we applied a fine-tuned \textit{BERT}-based sentiment classifier~\cite{Devlin.2019bert}, trained and validated on the annotated \textit{Stack Overflow} corpus by Novielli et al.~\cite{novielli2018stackoverflow-gold} using 10-fold cross-validation. Posts were classified into negative, neutral, or positive polarity following the emotion mapping in~\cite{noviellicross20}. The model achieved an average F1-macro score of 0.89 and an F1-micro score of 0.89. Implementation used \textit{Transformers~4.44} with default parameters and the \textit{BERT-base-uncased} model. 

\subsubsection{Variables}

\begin{table}[ht]
    \centering
    \caption{Variables and abbreviations used in the analysis.}
    \label{tab:variables-overview}
    \footnotesize
    \setlength{\tabcolsep}{3pt}
    \begin{tabularx}{\columnwidth}{Xl}
        \toprule
        \textbf{Variable (Abbreviation)} & \textbf{Scale / Range} \\ 
        \midrule
        \multicolumn{2}{l}{\textbf{Usefulness Metric}} \\ 
        Usefulness score ($U\textsubscript{ps}$) & Interval $\mathbb{R}$ \\ 
        \midrule
        \multicolumn{2}{l}{\textbf{Content Features}} \\ 
        Sentiment polarity ($E\textsubscript{pol}$) & Nominal $\{-1,0,1\}$ \\ 
        Code blocks / links ($E\textsubscript{cb}$, $E\textsubscript{ln}$) & Ratio $\mathbb{N}_0$ \\ 
        Has code / links ($E\textsubscript{has\_cb}$, $E\textsubscript{has\_ln}$) & Nominal $\{0,1\}$ \\ 
        Paragraphs / words / sentences ($E\textsubscript{pg}$, $E\textsubscript{wd}$, $E\textsubscript{st}$) & Ratio $\mathbb{N}_0$ \\ 
        Words per sentence ($E\textsubscript{wps}$) & Ratio $\mathbb{Q}_0^+$ \\ 
        Readability ($E\textsubscript{read}$) & Interval $(-\infty,206.8]$ \\ 
        Lexical diversity ($E\textsubscript{lex}$) & Ratio $\mathbb{Q}_0^+$ \\ 
        \midrule
        \multicolumn{2}{l}{\textbf{Relevance Features}} \\ 
        Similarity to question / peers ($E\textsubscript{simQ}$, $E\textsubscript{simE}$) & Ratio $[0,1]$ \\ 
        \midrule
        \multicolumn{2}{l}{\textbf{Temporal Features}} \\ 
        Time since question ($E\textsubscript{time}$) & Ratio (sec) \\ 
        Time of day ($E\textsubscript{tod}$) & Interval $[0,24)$ \\ 
        Creation date ($E\textsubscript{cd}$) & Interval (timestamp) \\ 
        Answers after ($E\textsubscript{ans\_after}$) & Ratio $\mathbb{N}_0$ \\ 
        \midrule
        \multicolumn{2}{l}{\textbf{Explainer Features}} \\ 
        Author reputation ($E\textsubscript{rep}$) & Ratio $\mathbb{N}_0$ \\ 
        Author badges (bronze/silver/gold) ($E\textsubscript{badges}$) & Ratio $\mathbb{N}_0$ \\ 
        Editor reputation / badges ($E\textsubscript{rep\_edit}$, $E\textsubscript{badges\_edit}$) & Ratio $\mathbb{N}_0$ \\ 
        \bottomrule
    \end{tabularx}
\end{table}

We distinguished between \textit{raw variables} (directly retrieved from the Stack Overflow API) and \textit{computed variables} (derived from text processing and metadata). 
Raw variables include votes, timestamps, and user reputation; computed variables capture readability, lexical diversity, and textual similarity.  
Table~\ref{tab:variables-overview} summarizes the analyzed variable groups.

$U\textsubscript{ps}$ quantifies normalized usefulness (relative score).  
$E\textsubscript{pol}$ represents sentiment polarity detected by a fine-tuned BERT model.  
Content measures ($E\textsubscript{cb}$–$E\textsubscript{lex}$) capture structure and readability.  
Relevance metrics ($E\textsubscript{simQ}$, $E\textsubscript{simE}$) assess textual overlap.  
Temporal variables ($E\textsubscript{time}$–$E\textsubscript{ans\_after}$) model posting dynamics.  
Explainer features ($E\textsubscript{rep}$–$E\textsubscript{badges\_edit}$) reflect author reputation and expertise.

Collectively, these variables operationalize structural, temporal, and social facets of \textit{developer explainability}.

\subsubsection{Hypotheses and Statistical Testing}

For each research question, we formulated corresponding null hypotheses (\(H_0\)) assuming no relationship between usefulness and the investigated feature group (sentiment, content, timing, author). 
Table~\ref{tab:hypothesen-uebersicht} summarizes all hypotheses and their associated variables. 
Type~I error inflation from multiple testing was mitigated using Bonferroni correction (\(\alpha_{\text{adj}} = \alpha / k\)), with a base significance level of \(\alpha = 0.05\).

\begin{table}[ht]
    \centering
    \caption{Overview of tested hypotheses and associated variables.}
    \label{tab:hypothesen-uebersicht}
    \scriptsize
    \begin{tabularx}{\textwidth}{lXp{4cm}}
        \toprule
        \textbf{Hypo.} & \textbf{Description (There is no relation...)} & \textbf{Used Variables} \\ 
        \midrule
        $H1_0$ & ...between usefulness and sentiment polarity of the explanation. & $U\textsubscript{ps}$, $E\textsubscript{pol}$ \\ 
        $H2_0$ & ...between usefulness and the structural content of the explanation. &  \\ 
        $H2.1_0$ & ...and the presence or number of embedded resources (code or links). & $E\textsubscript{cb}$, $E\textsubscript{ln}$, $E\textsubscript{has\_cb}$, $E\textsubscript{has\_ln}$ \\ 
        $H2.2_0$ & ...and textual length or density measures. & $E\textsubscript{pg}$, $E\textsubscript{wd}$, $E\textsubscript{st}$, $E\textsubscript{wps}$ \\ 
        $H2.3_0$ & ...and textual complexity and expressiveness. & $E\textsubscript{read}$, $E\textsubscript{lex}$ \\ 
        $H2.4_0$ & ...and similarity to the question or peer explanations. & $E\textsubscript{simQ}$, $E\textsubscript{simE}$ \\ 
        $H3_0$ & ...between usefulness and the timing of the explanation. & $E\textsubscript{tod}$, $E\textsubscript{time}$, $E\textsubscript{ans\_after}$, $E\textsubscript{cd}$ \\ 
        $H4_0$ & ...between usefulness and characteristics of the explainer. & $E\textsubscript{rep}$, $E\textsubscript{badges}$, $E\textsubscript{rep\_edit}$, $E\textsubscript{badges\_edit}$ \\ 
        \bottomrule
    \end{tabularx}
\end{table}

\subsubsection{Correlation Analysis}

Non-parametric correlations were used to identify associations between explanation features and usefulness:

\begin{itemize}
    \item \textbf{Spearman’s} $\rho$ for ordinal and ratio variables (e.g., length, reputation).  
    \item \textbf{Point-biserial} $r_{pb}$ for binary variables (e.g., code or link presence).  
    \item \textbf{Eta coefficient} $\eta$ for nominal variables (e.g., sentiment polarity).  
\end{itemize}

All coefficients were calculated with two-tailed significance testing and Bonferroni correction at the hypothesis level. 
Results were interpreted in terms of communicative effectiveness, whether structural, contextual, or author-related features consistently aligned with higher perceived usefulness in developer explanations.

\section{Results}
\label{sec:results}


\subsection{Data Overview}

We analyzed 3{,}323 \textit{Stack Overflow} questions and 59{,}398 answers (\(\sim\)17.9 answers/question on average) across 2{,}348 tags, contributed by 9{,}146 unique users (85{,}733 including commenters). The average relative usefulness score per answer was \(0.0559\). Activity peaked mid-week (most on Wednesdays, fewest on Saturdays).

\subsection{Correlation Analysis}
\label{sec:ergebnisse-der-korrelationsanalyse}

We examined associations between explanation features and their relative usefulness using non-parametric correlations (Spearman’s~$\rho$, Point-biserial~$r_{pb}$, and Eta~$\eta$; see replication package for details).  
All coefficients were computed two-tailed with Bonferroni correction per hypothesis group.

Table~\ref{tab:spearman_results} summarizes the results for continuous and ordinal predictors, while Table~\ref{tab:binary_nominal_results} reports correlations for binary and nominal variables.  
Together, these tables provide a comprehensive overview of how structural, temporal, and author-related features relate to perceived usefulness.

\textbf{Overall trends.}
Content-related features, such as the number of code blocks, links, and textual length, show small but consistent positive correlations with usefulness.  
In contrast, readability and lexical diversity exhibit negligible effects, indicating that linguistic complexity has little influence on perceived helpfulness.  
Timing variables show the strongest relationships: earlier answers (shorter delay since question) correlate clearly with higher usefulness.  
Among author-related attributes, reputation and badge counts correlate moderately positively, whereas last-editor metrics are negligible.  
Finally, sentiment polarity (negative/neutral/positive) shows virtually no association with usefulness, underscoring that emotional tone is irrelevant in this task-oriented setting.

\begin{table*}[ht]
    \centering
    \caption{Spearman correlation results (dependent variable: relative usefulness score $U\textsubscript{ps}$).}
    \label{tab:spearman_results}
    \scriptsize
    \setlength{\tabcolsep}{4pt}
    \renewcommand{\arraystretch}{0.9}
    \begin{tabularx}{\textwidth}{lXrrrrr}
        \toprule
        \textbf{Hyp.} & \textbf{Variable} & \textbf{$\rho$} & \textbf{Effect} & \textbf{Sig.} & \textbf{N} \\
        \midrule
        $H2.1.1.1_{0}$ & \textit{num code blocks} & 0.26 & Small & $<.0001$ & 59{,}398 \\
        $H2.1.2.1_{0}$ & \textit{num links} & 0.23 & Small & $<.0001$ & 59{,}398 \\
        $H2.2.1_{0}$ & \textit{num paragraphs} & 0.24 & Small & $<.0001$ & 59{,}398 \\
        $H2.2.2_{0}$ & \textit{num words} & 0.19 & Small & $<.0001$ & 59{,}398 \\
        $H2.2.3_{0}$ & \textit{num sentences} & 0.17 & Small & $<.0001$ & 59{,}398 \\
        $H2.2.4_{0}$ & \textit{words per sentence} & 0.10 & None & $<.0001$ & 59{,}398 \\
        $H2.3.1_{0}$ & \textit{readability} & -0.06 & None & $<.0001$ & 11{,}700 \\
        $H2.3.2_{0}$ & \textit{lexical diversity} & 0.03 & None & $<.0001$ & 11{,}700 \\
        $H2.4.1_{0}$ & \textit{sim. to question} & 0.09 & None & $<.0001$ & 59{,}398 \\
        $H2.4.2_{0}$ & \textit{sim. to peers} & 0.16 & Small & $0.0058$ & 59{,}321 \\ 
        \midrule
        $H3.1_{0}$ & \textit{time of day} & 0.01 & None & $<.0001$ & 59{,}398 \\
        $H3.2_{0}$ & \textit{time since question} & -0.50 & Strong & $<.0001$ & 59{,}398 \\
        $H3.3_{0}$ & \textit{answers after} & 0.12 & Small & $0.0046$ & 59{,}398 \\ 
        $H3.4_{0}$ & \textit{creation date} & -0.26 & Small & $<.0001$ & 59{,}398 \\ 
        \midrule
        $H4.1.1_{0}$ & \textit{owner reputation} & 0.41 & Moderate & $<.0001$ & 58{,}979 \\
        $H4.1.2_{0}$ & \textit{owner badges (bronze)} & 0.27 & Small & $<.0001$ & 58{,}979 \\
        $H4.1.3_{0}$ & \textit{owner badges (silver)} & 0.34 & Moderate & $<.0001$ & 58{,}979 \\
        $H4.1.4_{0}$ & \textit{owner badges (gold)} & 0.31 & Moderate & $<.0001$ & 58{,}979 \\
        $H4.2.1_{0}$ & \textit{editor reputation} & 0.05 & None & $<.0001$ & 31{,}315 \\
        $H4.2.2_{0}$ & \textit{editor badges (bronze)} & 0.04 & None & $<.0001$ & 31{,}315 \\
        $H4.2.3_{0}$ & \textit{editor badges (silver)} & 0.04 & None & $<.0001$ & 31{,}315 \\
        $H4.2.4_{0}$ & \textit{editor badges (gold)} & 0.00 & None & $0.47$ & 31{,}315 \\
        \bottomrule
    \end{tabularx}
\end{table*}

\begin{table}[ht]
    \centering
    \caption{Binary/nominal features vs.\ usefulness ($U\textsubscript{ps}$): point-biserial and eta results.}
    \label{tab:binary_nominal_results}
    \scriptsize
    \setlength{\tabcolsep}{3pt}
    \renewcommand{\arraystretch}{0.9}
    \begin{tabularx}{\columnwidth}{l l X r r r r}
        \toprule
        \textbf{Hyp.} & \textbf{Method} & \textbf{Independent variable} & \textbf{Coef.} & \textbf{Effect} & \textbf{Sig.} & \textbf{N} \\
        \midrule
        $H2.1.1.2_{0}$ & Point-biserial & has code blocks & 0.08 & None & $<.0001$ & 59{,}398 \\
        $H2.1.2.2_{0}$ & Point-biserial & has links & 0.19 & Small & $<.0001$ & 59{,}398 \\
        $H1_{0}$       & Eta ($\eta$)    & sentiment polarity & 0.0009 & None & $<.0001$ & 59{,}398 \\
        \bottomrule
    \end{tabularx}
\end{table}

\setlength{\shadowsize}{2pt}
\noindent
\shadowbox{
\begin{minipage}{0.94\columnwidth}
\textbf{Finding:} Usefulness increases with the inclusion of code and links, longer and earlier answers, and higher author reputation. Linguistic complexity, post-editing, and sentiment show no meaningful association.
\end{minipage}
}

\subsection{Hypotheses Evaluation}
\label{sec:hyp_eval}

Table~\ref{tab:hypothesis_results} summarizes the hypothesis testing outcomes. Bonferroni correction~\cite{weisstein2004bonferroni} was applied to mitigate Type~I error inflation.  
Due to the large sample size, even weak correlations reached statistical significance; thus, interpretation focuses on effect strength rather than $p$-values alone.

\begin{table*}[ht]
    \centering
    \caption{Hypothesis testing with Bonferroni-adjusted thresholds.}
    \label{tab:hypothesis_results}
    \scriptsize
    \setlength{\tabcolsep}{6pt}
    \renewcommand{\arraystretch}{0.9}
    \begin{tabularx}{\textwidth}{l r r X l}
        \toprule
        \textbf{H$_0$} & \textbf{\# Sub-Hyp.} & \textbf{Threshold} & \textbf{Rejected Sub-Hypotheses} & \textbf{Result} \\
        \midrule
        $H1_0$ & 1 & $0.05$ & – & rejected \\
        $H2_0$ & 12 & $0.0042$ & $H2.1_0$–$H2.4_0$ & Rejected \\
        $H3_0$ & 4 & $0.0125$ & $H3.1_0$–$H3.4_0$ & Rejected \\
        $H4_0$ & 8 & $0.0063$ & $H4.1_0$, $H4.2_0$ & Rejected \\
        \bottomrule
    \end{tabularx}
\end{table*}

\noindent
$H1_0$ (sentiment polarity) was rejected, but the effect size is negligible, confirming that emotional tone does not affect perceived usefulness.  
$H2_0$ (content), $H3_0$ (timing), and $H4_0$ (author) were also rejected, indicating that structural richness, timeliness, and contributor reputation significantly influence perceived usefulness.  
These findings reinforce that structural and contextual features are stronger predictors of communicative effectiveness than linguistic tone or post-editing metadata.

\section{Discussion}
\label{sec:discussion}

This section interprets the results and answers the research questions.

\subsection{Answering the Research Questions}
\label{sec:beantworten-der-forschungsfragen}

\textbf{RQ1: How does the usefulness of an explanation relate to its sentiment polarity?}  
Sentiment showed no meaningful association with usefulness. The negligible effect supports the view that developer communication is primarily functional and goal-oriented, where clarity and factual accuracy outweigh emotional tone.

\textbf{RQ2: How does usefulness relate to content features such as code inclusion, text length, and readability?}  
Content-related factors, particularly the number of code blocks, links, and explanation length, showed small but consistent positive correlations with usefulness. This suggests that richer, example-driven explanations tend to be perceived as more helpful, consistent with prior findings on answer quality and with principles of clarity and completeness in requirements communication.

\textbf{RQ3: How does timing (e.g., response order and posting delay) influence usefulness perception?}  
Temporal factors show the strongest effects. Earlier answers receive higher usefulness scores, reflecting visibility advantages and the importance of timely feedback. In RE terms, this parallels how early clarification reduces uncertainty and guides subsequent discussion.

\textbf{RQ4: How do author characteristics (e.g., reputation, badges) affect perceived usefulness?}  
Author reputation and badge counts are moderately correlated with usefulness, while editor attributes have negligible influence. This suggests that perceived credibility affects evaluations independently of content, pointing to a potential reputation bias similar to authority effects in collaborative RE activities.

\medskip
Overall, structural and contextual factors best explain perceived usefulness. Timely, detailed, and reputable explanations tend to be valued more, suggesting that in developer-to-developer and RE-like communication, usefulness is driven by informational quality and context rather than affective tone.

\subsection{Reflection on Findings and Implications}
\label{sec:interpretation}

Our findings show that in developer communication, and by extension in requirements engineering (RE), perceived usefulness is primarily shaped by structural and contextual factors rather than linguistic tone. Content richness, timing, and author reputation are associated with higher perceived usefulness, whereas emotional style contributes little.

This challenges assumptions that sentiment meaningfully affects quality perceptions. In task-oriented contexts such as RE, communication serves knowledge transfer, and clarity, completeness, and evidence matter most.

Content-related attributes such as code blocks, links, and explanation length were associated with modest but consistent increases in usefulness, supporting prior work showing that detailed, example-based explanations improve comprehension~\cite{calefato16stack,omondiagbe2019features}. At the same time, such structural richness can also amplify fluency or verbosity effects, where well-written and confidently phrased explanations may be perceived as helpful even when their technical grounding is weak. This aligns with recent evidence that LLM-generated answers can be verbose and well-articulated, and that user study participants may still prefer such answers even when they contain incorrect information, indicating a fluency-driven trust risk~\cite{kabir20242024SOobsolet}. Complementing this concern, \cite{obaidi2025explainability} reports that users can often recognize LLM-formulated explanations and tend to prefer manually written ones when LLM outputs appear less correct or less aligned with the underlying facts, underscoring that perceived clarity should not be conflated with correctness. Timing was the strongest predictor: earlier answers were rated more useful, reflecting both visibility effects and the value of prompt clarification, paralleling how early feedback in RE reduces ambiguity and supports shared understanding.


Author reputation showed moderate positive effects, indicating possible credibility bias~\cite{wang2021reputation}. In RE practice, this underlines the need for transparent authorship and balanced peer review to avoid overreliance on perceived expertise when judging rationale quality.

From a practical perspective, the results suggest two complementary implications. For developers and RE practitioners, useful explanations tend to be detailed, contextually grounded, and supported by concrete artifacts (e.g., code, links, measurable acceptance criteria, scenarios). Accordingly, RE communication and documentation should capture clarification and rationale early and in an example-based form, while reviews and tools should reduce visibility and authority bias by prioritizing well-structured content over popularity or posting time.

For teaching explainable AI, the findings motivate training students to ground explanations in evidence and traceable assumptions (e.g., examples or snippets) and to explicitly reflect on fluency and reputation cues, so that well-written explanations are not automatically treated as correct or trustworthy.


In sum, effective explainability emerges as a communicative competence grounded in structure and context, not style. Developers who explain early and substantively contribute most to collective understanding and decision quality in software projects.

\subsection{Threats to Validity}
\label{sec:validiteat}

We discuss threats to validity following Wohlin et al.~\cite{wohlin2024experimentation}.

\noindent\textbf{Construct Validity.}
Usefulness was approximated through normalized engagement metrics (relative score), which mitigate but do not eliminate visibility or popularity effects~\cite{anderson2012stackoverflow,wang2021reputation}. Upvotes were chosen as the most widely used proxy in prior Q\&A research, balancing availability and comparability across threads. However, votes do not measure correctness, comprehension, or ambiguity reduction directly; they primarily capture perceived usefulness under platform-specific biases. Sentiment was derived via automated classification and may misinterpret neutral or technical phrasing. Jaccard similarity, being surface-based, cannot fully capture semantic relatedness.

\noindent\textbf{Internal Validity.}
As the analysis is correlational, causal inferences cannot be made~\cite{10.1093/acprof:oso/9780199764044.003.0003}. Uncontrolled confounders such as topic difficulty~\cite{vasilescu2014stackoverflow}, author expertise, or community norms may influence results. Reputation might also act both as cause and effect of perceived usefulness.

\noindent\textbf{Conclusion Validity.}
Skewed sentiment class distributions could attenuate effects. Bonferroni correction reduced false positives but increased the chance of false negatives. Most significant effects were small, indicating limited practical impact despite statistical significance. Given the large sample size, even small effects reached statistical significance, emphasizing the need to interpret effect strength rather than p-values.

\noindent\textbf{External Validity.}
The dataset covers highly upvoted, English-language \textit{Stack Overflow} posts and may not generalize to other communities or less technical audiences. Platform dynamics may also affect applicability over time. Post-2022 shifts in developer workflows, including increased LLM use, can change answer style and user preferences and may therefore influence voting-based signals~\cite{kabir20242024SOobsolet}. The dataset focuses on high-quality Android-tagged questions ($\geq 50$ upvotes), which may bias the sample toward established topics and may not reflect other domains such as security or ML. Stack Overflow captures developer-to-developer Q\&A rather than stakeholder-facing RE communication, but similar mechanisms such as visibility effects, authority cues, and timing can shape which explanations receive attention in RE discussions. Finally, generalizability across replications may be sensitive to analytic choices, including preprocessing, sentiment models, and the use of bivariate correlations without controlling for confounders.

\section{Conclusion and Future Work}
\label{sec:conclusion}

This study analyzed 59{,}398 \textit{Stack Overflow} explanations to identify which factors shape their perceived usefulness. Results indicate that usefulness depends primarily on structural and contextual qualities: explanations that are timely, detailed, and supported by code or links are rated most helpful. In contrast, sentiment has no meaningful influence, suggesting that clarity and substance outweigh linguistic tone in developer communication. Explainer reputation shows a moderate effect, implying that credibility cues partially drive perceived quality—a potential source of reputation bias also relevant for requirements evaluation. Overall, the findings underline that effective explainability in both online and RE contexts is grounded in structure and context rather than affective style. Clear, complete, and well-timed explanations strengthen shared understanding and decision quality in software projects. 

Future work should examine the generalizability of these results across platforms with different audiences (e.g., \textit{Reddit}, \textit{Quora}) and complement quantitative analyses with qualitative studies to better understand why certain explanations are perceived as useful. Further studies should test explicit mappings from explanation features to RE practices, e.g., linking code-supported explanations to requirements-by-example (acceptance criteria) and timely explanations to early clarification in elicitation and validation. In addition, refining textual features (e.g., specificity, factual accuracy, confidence) and running controlled experiments (e.g., anonymized authorship) could clarify whether reputation effects reflect actual quality or perception bias and support fairer surfacing of useful explanations.

\section*{Acknowledgment}
This work was funded by the Deutsche Forschungsgemeinschaft (DFG, German Research Foundation) under Grant No.: 470146331, project softXplain (2022-2025).
\subsubsection*{Data Availability Statements}
{\fontsize{9pt}{11pt}\selectfont All data of our study is publicly available at \href{https://zenodo.org/records/17988272}{Zenodo}}~\cite{obaidi2026datasetSO}.

%

\bibliographystyle{splncs04}
\bibliography{references} 

\end{document}